\documentclass[12pt]{article}
\usepackage{graphicx}
\newcommand{\beq}{\begin{equation}}
\newcommand{\eeq}{\end{equation}}
\newcommand{\bea}{\begin{eqnarray}}
\newcommand{\eea}{\end{eqnarray}}

\def\fun#1#2{\lower3.6pt\vbox{\baselineskip0pt\lineskip.9pt
  \ialign{$\mathsurround=0pt#1\hfil##\hfil$\crcr#2\crcr\sim\crcr}}}
\begin{document}
\begin{titlepage}
\begin{flushleft}
%       \hfill                      {\tt hep-th/03010***}\\
       \hfill                       FIT HE - 03-05 \\
\end{flushleft}
\vspace*{3mm}
\begin{center}
{\bf\LARGE Effective action and brane running\\ }
%{\bf\LARGE 
% for brane-world \\ }
\vspace*{5mm}

\bigskip

{\large Iver Brevik \footnote{\tt iver.h.brevik@mtf.ntnu.no}\\}
\vspace{2mm}
{
%\large 
Department of Energy and Process Engineering, Norwegian University of Science and Technology,
N-7491 Trondheim, Norway\\}
\vspace*{5mm}

{\large Kazuo Ghoroku\footnote{\tt gouroku@dontaku.fit.ac.jp}\\ }
\vspace*{2mm}
{
%\large 
${}^2$Fukuoka Institute of Technology, Wajiro, Higashi-ku}\\
{
%\large 
Fukuoka 811-0295, Japan\\}
\vspace*{5mm}

{\large Masanobu Yahiro \footnote{\tt yahiro@sci.u-ryukyu.ac.jp} \\}
\vspace{2mm}
{
%\large 
${}^4$Department of Physics and Earth Sciences, University of the Ryukyus,
Nishihara-chou, Okinawa 903-0213, Japan \\}

\vspace*{10mm}

\end{center}

\begin{abstract}
We address the renormalized effective action for a  Randall-Sundrum 
brane running in 5d bulk space. The running behavior of the brane
action is obtained by shifting the brane-position without changing
the background and the fluctuations.
After an appropriate renormalization, we obtain
an effective, low energy braneworld action, in which 
the effective 4d Planck mass is 
independent of the running-position. 
We address some implications of this effective action.

\end{abstract}
\end{titlepage}

\section{Introduction}
It is quite probable that our 4d world is formed 
according to a ten-dimensional superstring theory. 
In particular, an interesting
geometry $AdS_5\times S^5$ is obtained near the horizon of the stacked 
D3-branes in type IIB string theory.
And the string theory on this background  may describe
a four-dimensional conformal field theory (CFT), $\mathcal{N}=4$
SUSY Yang-Mills theory, 
which lives on the boundary \cite{M1,GKP1,W1,Poly1}. 

%\vspace{.2cm}
On the other hand, a thin three-brane (Randall-Sundrum brane)
embedded in $AdS_5$ space ~\cite{RS1,RS2} can be regarded as the boundary
pushed out to AdS bulk. Since the coordinate transverse to the brane
is considered as the energy scale of CFT on the boundary, 
the position of the brane represents the ultraviolet
cut-off scale of the CFT.

\vspace{.2cm}
Usually,
the brane action is expressed in terms of a tension  
 only, when  solving the equation
of motion for the braneworld. It has recently been pointed out that 
many higher derivative terms appear in the brane action when we make
a change of the brane position without changing the solutions for the
background and the Kaluza-Klein (KK) modes \cite{LMS,LR,Re,U}.
This procedure is called  brane running and is considered as an approximate
form of obtaining renormalization group flow.
This implies that the parameters of the brane-action at a different
position could be related through this renormalization group flow.
Inspired by this idea, we examine the running behavior of the brane action and
derive the effective braneworld action 
with an appropriate renormalization.

\vspace{.2cm}
We restrict ourselves to the case of pure gravity in the 5d bulk. It is
well known that
the zero mode is trapped on the brane and the 
usual 4d Newton's law can be observed on the brane.
Non-trapped, massive KK modes are also observed through a 
correction to  Newton's law, 
and this is also understood
from the 4d field theory based on the AdS/CFT correspondence \cite{DL,NOZ}. 
We address the effects of KK modes through the running brane-action, which
is obtained as a solution of the flow equations. But ultra-violet
divergences are generally included in the brane-action. This is
generally expected from the viewpoint of higher dimensional field theory,
which would correspond to  CFT on the brane,
since loop-corrections are in general included in the brane action 
\cite{GR,GGH,MOZ}.
Then the effective action is
obtained after renormalizing the running brane-action
in a consistent way via the AdS/CFT correspondence.
%%%%%%%%%%% 2nd modified
% As a result,
When we set the brane near the boundary of AdS$_5$, the higher derivative
terms are neglected and
%%%%%%%%%%%
we arrive at an low energy, effective braneworld action in which
the effective 4d Planck mass is independent of the flow-position. 
In other words,
the original Randall-Sundrum formulation, in which the brane action is
written in terms of tension only, is available at any flowing position as far as
 low energy theory is considered.
 
\vspace{.2cm}
In Section 2, the model used here is set, and the flow equation for the
tension is derived and solved in  Section 3. 
In Section 4, the flow equations for the brane action are derived
and solved. The effective brane action is derived by an appropriate
renormalization condition, and its meaning is  discussed.
In Section 5, the case of two parallel branes is discussed and the problem
of their distance is commented upon. %In section 6, the cosmological implications
%are given. 
Concluding remarks are given in the final section.

\section{Setting of the brane-world}

We start from the five-dimensional gravitational action. It is given in the
Einstein frame as\footnote{
Here we take the following definition, $R_{\nu\lambda\sigma}^{\mu}
=\partial_{\lambda}\Gamma_{\nu\sigma}^{\mu}-\cdots$, 
$R_{\nu\sigma}=R_{\nu\mu\sigma}^{\mu}$ and $\eta_{AB}=$diag$(-1,1,1,1,1)$. 
Five dimensional suffices are denoted by capital Latin and four
dimensional
ones by  Greek letters.
}
\beq
    S_5 = {1\over 2\kappa^2}\Bigg\{
      \int d^5X\sqrt{-G} (R -  2\Lambda)
          +2\int d^4x\sqrt{-g}K\Bigg\}, \label{action}
\eeq
where 
$K$ is  the extrinsic curvature on the boundary. 
The other ingredient is the brane action,
\beq
    S_{\rm b} = -{\tau}\int d^4x\sqrt{-g}, \label{baction}
\eeq
which is added to $S_5$, and the Einstein equation is obtained as
\beq
 R_{MN}-{1\over 2}g_{MN}R=\kappa^2T_{MN} \label{equation}
\eeq
where $\kappa^2T_{MN}=-(\Lambda+b\delta(y)\kappa^2\tau
\delta_{\mu}^M\delta_{\nu}^N) g_{MN}$ and $b=\sqrt{-g}/\sqrt{-G}$.
Here we solve the Einstein equation (\ref{equation}) with the following
metric,
%%%%%%%%%%%%%%%%%%%%%%%%%%%%%%
\beq
 ds^2= A^2(y)\left\{-dt^2+a_0^2(t)\gamma_{ij}(x^i)dx^{i}dx^{j}\right\}
           +dy^2  \, \label{metrica},
\eeq
where the coordinates parallel to the brane are denoted by $x^{\mu}=(t,x^i)$,
$y$ being the coordinate transverse to the brane. The position of the brane
is taken at $y=0$. We restrict our interest here to the case of a
Friedmann-Robertson-Walker type
(FRW) universe. Then, the three-dimensional metric $\gamma_{ij}$
is described in Cartesian coordinates as
\beq
  \gamma_{ij}=(1+k\delta_{mn}x^mx^n/4)^{-2}\delta_{ij},  \label{3metric}  
\eeq
where the parameter values $k=0, 1, -1$ correspond to a
 flat, closed, or open universe respectively.

\vspace{.3cm}
When  considering the metric (\ref{metrica}), we obtain 
the following reduced equations \cite{bre}:
\beq
  ({\dot{a_0}\over a_0})^2+{k\over a_0^2}=A'^2+{\Lambda\over 6}A^2
          =D,  \label{Einstein2}
\eeq
where $D$ is a constant being independent of $t$ and $y$. 
In view of the boundary condition at the brane position,
\beq
  A'(0)=-{\kappa^2\tau\over 6}A(0), \label{bound2}
\eeq
one gets
\beq
   D\equiv \lambda = \kappa^4\tau^2/36+\Lambda/6 , \label{4cos}
\eeq 
if  $A(0)=1$. Here we use the following notation, $\dot{}=d{}/dt$ and ${}'=d{}/dy$.
The normalization condition
$A(0)=1$ does not affect the generality of our discussion.

\vspace{.5cm}
Here $\lambda$ denotes the cosmological constant on the brane as seen from
the first equation of (\ref{Einstein2}),
\beq
 ({\dot{a_0}\over a_0})^2+{k\over a_0^2}=\lambda \ . \label{cc}
\eeq
While the tension of the brane is expressed by this $\lambda$ according to
(\ref{4cos}),
\beq
 \tau={6\mu\over \kappa^2}\sqrt{1+{\lambda\over \mu^2}}, 
 \quad \mu=\sqrt{-\frac{\Lambda}{6}}. \label{desitt-tau}
\eeq

\vspace{.5cm}
This result is obtained for a fixed brane-position at $y=0$, but
the position is arbitrary in the above discussion. In the context of
AdS/CFT correspondence, the coordinate $y$
has the meaning of an energy-scale in the field theory on the brane. 
We can obtain some information about renormalization group flow of the 
field
theory when the brane at different positions of $y$ are related by 
an appropriate 
principle. In the following, we study this problem by using the boundary conditions of the
field equations and by extending the brane action such that it 
can change with $y$.

\section{Flow of $\tau$}

In the previous section, $\tau$ in (\ref{desitt-tau}) is obtained in terms of
(\ref{baction}) and (\ref{bound2}) for the brane-position $y_0=0$. Here
we consider a shift the position from $y_0$ to $y_1=y_0+\epsilon$
without changing the bulk configuration $A(y)$. Then we obtain the
following boundary condition, which can be written in the same form 
as (\ref{bound2}),
\beq
   A'(y_1)=-{\kappa^2\tau(y_1)\over 6}A(y_1). 
\label{bound21}
\eeq
Here we have replaced $\tau$ by $ \tau(y)$, a function of $y$.
In other words, this shift describes the change of $\tau (y)$ with $y$ in accordance with the
change of $A(y)$ with $y$. This means that the parameter of the
brane action must run with $y$ so as to satisfy (\ref{bound21}).

Then for the shift of $y_0$ keeping the bulk configuration unchanged,
$\tau(y)$ is obtained for any value of $y$ as
\beq
  \tau(y)=-{6\over\kappa^2}{A'(y)\over A(y)}. 
\label{bound22}
\eeq

\vspace{.3cm}
This result can be also obtained by solving a differential equation that is
derived by expanding (\ref{bound21}) in a series of
$\epsilon$ when we take $y_1=y_0+\epsilon$. It is given as
\beq
 \tau'={\Lambda\over \kappa^2}
         +{\kappa^2\over 6}\tau^2.  \label{RG1}
\eeq
In getting this equation, we used the Einstein equations for $A$. It is easy
to show the equivalence of the solution of this equation and the one given
by (\ref{bound22}). This is because $A$ obeys 
the following linear equation,
\beq
  {A''\over A}+{\Lambda\over 6}=-{\kappa^2\tau\over 3}\delta(y),
\eeq 
whereby  $A''$ can be rewritten  in terms of $A$.

\vspace{.3cm}
We will consider some  explicit examples of flows mentioned above, in terms of 
typical brane-world solutions.

\subsection{Solutions for $\lambda=0$}

When $\lambda=0$, the solution is given by
\beq
 ds^2= e^{-2\mu |y|}(-dt^2+\delta_{ij}dx^{i}dx^{j})
           +dy^2.  \, \label{metrica22}
\eeq
In this case 
 $A= e^{-\mu |y|}$, and we obtain from (\ref{bound22})
\beq
 \tau(y)={6\mu\over\kappa^2}. 
\label{tau0-lambda0}
\eeq
This  corresponds to a fixed point in the sense
of flow of $\tau$  as we should expect, i.e.,
\beq
 \tau'=0.
\eeq
This is also assured from (\ref{RG1}) by adopting the "initial"
value of $\tau$ as $\tau(0)={6\mu\over\kappa^2}$.

However, this does not mean that there is no $y$-dependent terms
in the brane action. As shown in (\ref{baction2}) in the next subsection,
the general form of the brane action could include many terms, but at present 
we see only the first term $\tau_0$ appearing in (\ref{baction2}).
Actually, we can see other $y$-dependent  terms
from the solution of the "RG" equation derived from the 
gravitational fluctuations. We will show this  in the next section.

\subsection{Solutions for $\lambda >0$ and $\Lambda<0$}

For $\lambda=0$, there was no special position other than
the position of the brane, $y_0$. 
But for $\lambda>0$  a horizon $y_H$, where
$A(y_H)=0$, appears.   We consider the region of $y_H>y_0$
hereafter.

When $\lambda >0$ and $\Lambda$ is negative, one has
\beq
 A(y)={\sqrt{\lambda}\over \mu}\sinh[\mu(y_H-|y|)],
          \label{desit0}
\eeq 
\beq
   \sinh(\mu y_H)=\mu/\sqrt{\lambda}.
               \label{const10}
\eeq
Here $y_H$ represents the horizon in the bulk AdS$_5$, and 
(\ref{const10}) is a condition for $A(0)=1$. However, we notice
that $A(0)$ might be arbitrary and $A(0)=1$ is at present chosen for simplicity.
 This point will be discussed below.
We get
\beq
 \tau(y)={6\mu\over\kappa^2}\coth(\mu[y_H-y]), \qquad
\eeq
which gives interesting information.
%It diverges at $y=y_H$, so we need some
%corrections from string theory near this point. 
In view of (\ref{const10}) we obtain
the following expansion of $\tau$ with respect to $\lambda$,
\bea
   &&\tau(y)={6\mu\over \kappa^2}\sqrt{1+{\lambda\over \mu^2 A^2}}
\nonumber
\\
&&~~~~~~~~~~={6\mu\over \kappa^2}+{3\over \kappa^2\mu}{\lambda\over A^2}+
               O(\lambda^2). \label{desitt}
\eea
The first line of the above equation is written as
\beq
  ({\kappa^2\tau\over 6})^2=\mu^2+{\lambda\over A^2},
\eeq
which represents the extended form of the parameter relation obtained at 
$y=0$, $({\kappa^2\tau\over 6})^2=\mu^2+{\lambda}$,
to that holding at arbitrary $y$.

As for the expansion in the second line, we suggest that  
the origin of the
$\lambda$-dependent terms is as follows. Consider the following form of
brane action including curvature terms,
\beq
 \tilde{S}_{\rm b} = -\int d^4x\sqrt{-g}
      \left(\tau_0+\tau_1 R
      +\tau_2\left(R_{\mu\nu}R^{\mu\nu}-{1\over 3}R^2\right)
      +\tilde{\tau}_2R^2+\cdots\right). 
   \label{baction2}
\eeq
Then we find the following form of $\tau$,
\beq
    \tau =\tau_0+{\tau_1}{6\lambda\over A^2}
             %-\tau_2{3\lambda^2\over A^4}
+ O(\lambda^3) ,
\eeq
for the de Sitter brane.
For the case of Minkowski brane, $\tau=\tau_0$, as stated above.
In the case of $\lambda>0$, $\tau$ should be constructed by many
curvature terms in general, if they were prepared in the brane action.
If $\tau_i(y), i\geq 1$ does not include $\lambda$, we find 
\beq
 \tau_0={6\mu\over \kappa^2}\, , \quad \tau_1=\frac{1}{2\kappa^2\mu}, 
 %\quad \tau_2=1/(4\kappa^2\mu^3), \quad \tilde{\tau}_2=0, 
\quad \cdots.
\label{sol-desitter}
\eeq 
We should notice that the first term is consistent with the one
given in the previous sub-section. However,
it turns out that the second term depends on $y$, as is shown in the next section.
Then $\tau_i$ themselves must depend on $\lambda$, and we will
need an infinite series of curvatures in the brane
action to allow for the shift of  brane position without changing the
background configuration. This is seen more explicitly in the next section
when the same condition is imposed on equations of the field-fluctuations.

For the cases of other brane solutions, we give a brief comment in the 
final section.

\vspace{.5cm}

\section{Flow of $\tau_i$, $i\ge 1$, and effective brane action}

Consider the perturbed metric $h_{ij}$ in the form
\beq
 ds^2= A^2(y)(-dt^2+a_0^2(t)[\gamma_{ij}(x^i)+h_{ij}(y,x^{\mu})]dx^{i}dx^{j})
           +dy^2  \,. \label{metricape}
\eeq
We are interested in the traceless transverse
component, which represents the graviton on the brane, of the perturbation.
It is projected out by the conditions, $h_i^i=0$ and
$\nabla_i h^{ij}=0$, where $\nabla_i$ denotes the covariant derivative
with respect to the three-metric $\gamma_{ij}$ which is used to raise
and lower the three-indices $i,j$. The transverse and traceless part
is denoted by $h$ hereafter for simplicity.

%\subsection{$\lambda >0$ case}
\vspace{.5cm}
We first consider the simple case where $\Lambda<0$ 
and $\gamma_{ij}(x^i)=\delta_{ij}$ .  Then the transverse
and traceless part $h$ is projected out by
$\partial_ih^{ij}=0$
and $h^i_i=0$, where $\delta_{ij}$ is used to raise and lower the indices
${ij}$. One arrives at the following linearized equation of $h$ in terms of
the five (four) dimensional covariant derivative $\nabla^2_5=\nabla_M\nabla^M$
($\nabla^2_4=\nabla_{\mu}\nabla^{\mu}$):
\beq
 {1\over 2\kappa^2}\nabla^2_5 h=\left(\tau_1{\nabla^2_4\over A(0)^2}
         -\tau_2{\nabla^4_4\over A(0)^4}+\cdots \right)h\delta(y).  \label{scalar}
\eeq
We notice that $\tau_0$ is not included on the right hand side since
it disappears due to the classical Einstein equation. And the right hand
side is expressed exactly for $\lambda=0$. When $\lambda$ is finite, $\tau_i$
should be replaced by $\bar{\tau}_i$ which include the terms coming from
higher curvature terms like $R^{i+n} (n\geq 1)$. In this sense, we can not
see the pure $\tau_i$ for the case of $\lambda>0$. For a while, we denote $\tau_i$
by the same notation for $\lambda=0$ and $\lambda>0$.

\vspace{.5cm}
Eq.~(\ref{scalar}) is rewritten as follows by the following 
factorization of $h$,
\beq
 h=\int dm \phi_m(t,x^i)\Phi(m,y) \, , \label{eigenex}
\eeq
where the mass $m$ is defined by
\beq
  \ddot{\phi}_m+3{\dot{a_0}\over a_0}\dot{\phi}_m
           +{-\partial_i^2\over a_0^2}\phi_m=-m^2\phi_m , \label{masseig}
\eeq
then the equation for $\Phi(m,y)$ is obtained as 
\beq
  {\Phi}''+4{A'\over A}{\Phi}'
           +{m^2\over A^2}\Phi=2\kappa^2\left(\tau_1{m^2\over A(0)^2}
         -\tau_2{m^4\over A(0)^4}+\cdots \right)\Phi(0)\delta(y) . \label{warpgr}
\eeq
In solving (\ref{warpgr}), we need the following boundary condition,
\beq
{\Phi'(0)\over \Phi(0)}={\kappa^2}(\tau_1{m^2\over A^2(0)}
-\tau_2{m^4\over A^4(0)}+\cdots).
\eeq
Then this boundary condition is different from the one used before \cite{bre},
where we used $\tau_i=0, i\geq 1$. The new boundary condition affects only
 the KK modes, but it does not give any effect on the zero mode. Then the
results for the localization of graviton, the zero mode,
are not changed even if we consider the above new boundary conditions.

\vspace{.5cm}
By shifting the position of brane, we obtain flow equations for $\tau_i$ as
above by defining $\tau_T=-\tau_1{m^2\over A^2}+\tau_2{m^4\over A^4}\cdots$,
\beq
  \tau_T'=-4{A'\over A}\tau_T+\kappa^2\tau_T^2+{m^2\over \kappa^2A^2}~~,
   \label{running2}
\eeq
>From this, flow equations for $\tau_i$ are obtained as,
\beq
  \tau_1'=-2{A'\over A}\tau_1-{1\over \kappa^2}~~,
 \quad \tau_2'={\kappa^2}\tau_1^2~~ , 
   \label{running}
\eeq
\beq
  \tau_3'=-2\kappa^2\tau_1\tau_2+2{A'\over A}\tau_3\, ,\quad \cdots~~ \quad .
   \label{running3}
\eeq

\subsection{The case $\lambda =0$ }

In this case, $\tau_1$ is obtained as,
\beq
 \tau_1={1\over 2\mu\kappa^2}+c_1e^{2\mu y},
\eeq
with an integration constant $c_1$.
For AdS$_5$ background, $\tau$ is independent of $y$ as seen in the 
previous section. So one idea to
determine $c_1$ is to demand that $\tau_1$ is also $y$-independent.
%In this case, we expect that parameters are not running from
%the viewpoint of AdS/CFT correspondence as seen in the previous section
%for $\tau_0$. As for $\tau_1$, 
This is realized by choosing  $c_1=0$.
%\beq
%  \tau_1(0)={1\over 2\mu\kappa^2}
%\eeq
%For this condition, we can see $\tau_1'=0$ and $\tau_1(y)=\tau_1(0)$.
However, $\tau_2$ should be running for the above setting of $\tau_1$
since the second equation of (\ref{running}) is written as
\beq
\tau_2'={1\over 4\mu^2\kappa^2}~~,
\eeq
and this is solved as
\beq
 \tau_2={1\over 4\mu^2\kappa^2}y+c_2,  \label{tau2-0}
\eeq
where $c_2$ is a constant. 
This term is interpreted as the anomaly produced by  one-loop 
 CFT,  from the viewpoint of AdS/CFT correspondence. In this sense, there
is no $y$-independent solution of $\tau_i, i\ge 1$. In other words, there
is no fixed point in the parameter space.

\vspace{.5cm}
Further, we obtain $\tau_3$ as
\beq
 \tau_3=c_3e^{-2\mu y}-{1\over 8\mu^4\kappa^2}y+\tilde{c}_2
   \label{tau3}
\eeq
where $\tilde{c}_2=-c_2/(2\mu^2)+1/(16\mu\kappa^2)$.
In the context of the renormalization group, this corresponds to an
irrelevant term. But it is running, as  it has received a quantum
correction from CFT similarly to $\tau_2$. 
We can see the running of other irrelevant terms
as well as $\tau_3$. Then the flow equations given here
would be valid as  renormalization group equations in the region
where the irrelevant terms vanish or are suppressed. 

The first term of $\tau_3$ becomes
%%%%%%%%%%%%%%%%%%%%% modified 2nd
small when the brane position $y_0$, which is set as $y_0=0$ in
equation (\ref{tau3}), is taken 
near the boundary, $y_0=-\infty$, since it is written as $c_3e^{-2\mu (y-y_0)}$
in this case
\footnote{When the brane position is taken at $y=y_0$, the warp factor is
written as $A(y)=A(y_0)e^{-\mu|y-y_0|}$ and $A(y_0)=e^{-\mu y_0}$. 
And the brane runs
from $y_0$ to the extended space direction $y>y_0$ according to the same
flow equations as given here.
}.
So we can neglect this term near the boundary. 
%%%%%%%%%%%%%%%%%%%%%% modified 2nd
The remaining terms
could be suppressed under the condition that the above flow equations
are valid for small $m^2/\mu^2$ or for small mass of the KK modes
compared to $\mu$. Then we can neglect the irrelevant terms for $m^2/\mu^2<<1$.
%%%%%%%%% Added
Even if $\tau_i, i\geq 3$ were remaining  finite, 
these irrelevant terms would vanish
in the action due to the warp factor $A^{-2i+4}$ in the limit of
$y_0\to -\infty$. In this sense, the effective action is well defined
near the boundary in terms of the relevant terms. Hereafter, we consider
the braneworld including only low frequency KK modes of $m^2/\mu^2<<1$
to remove the irrelevant terms in the effective braneworld action.

\vspace{.5cm}
Then we estimate the effective brane action $S^{\rm eff}_b$ from the
viewpoint of AdS/CFT correspondence neglecting the irrelevant terms. It
can be written as \cite{Gidd,GKa,HeSk,GY}
\beq
 S^{\rm eff}_b={1\over 2}\tilde{S}_b + \ln Z_5(g)
           = {1\over 2}\tilde{S}_b + S_{\rm CT}+S_{\rm CFT}
  \label{effaction}
\eeq
\beq
  Z_5(g)=\int_{G|_{y=0}=g} DG D\psi e^{iS_5},   \label{effective}
\eeq
where $\tilde{S}_b$ and $S_5$ are given in the previous section. 
The fields other than the metric contained in $S_5$ are denoted by $\psi$.
In the case of AdS$_5$, we can set  $\tau_0=\tau$, then
$\tilde{S}_b$ can be given by solving the above flow equation for other
$\tau_i$. The term $\ln Z_5(g)$ is replaced by $S_{\rm CT}$ (the counter term)
and $S_{\rm CFT}$ (cut-off CFT), which interacts with
the gravity in $\tilde{S}_b$ and provides the quantum correction
to $\tau_2$. The term $S_{\rm CT}$ is given as,
\beq
 {S}_{\rm CT} = \int d^4x\sqrt{-g}
      \left({3\mu\over \kappa^2}+{1\over 4\mu\kappa^2}R
      +{\ln(\mu z)\over 8\mu^3\kappa^2}\left(R_{\mu\nu}R^{\mu\nu}-{1\over 3}R^2\right)
      \right),
   \label{bactionCT}
\eeq
where $\mu y=\ln(\mu z)$.
%Then we obtain noticing $\mu y=\ln(\mu z)$, 
%\beq
% S^{\rm eff}_b = -\int d^4x\sqrt{-g}
%       {\tilde{c}_2\over 2}\left(R_{\mu\nu}R^{\mu\nu}-{1\over 3}R^2\right)
%           +S_{\rm CFT} .
%  \label{effactionAdS}
%\eeq
%\vspace{.5cm}
%The result (\ref{effactionAdS}) implies that there is no Einstein
%term in the effective brane action. Then the observer on the brane could not
%find the usual Newton law of potential $1/r$
%in spite of the localization of the graviton on the brane. This is the result
%of the cancellation of the localized zero mode of the bulk graviton and the
%graviton given in the brane action.
%Instead of $1/r$ potential, one might observe 
%the gravitational
%potential $1/r^3$ or the Wely gravity which would be considered as the 
%correction coming from CFT or the KK modes in the bulk.
%Then this brane does not describe our universe. 

\vspace{.5cm}
As for $\tilde{S}_b$, we consider the boundary condition of the
usual brane model, $\tau_i(0)=0, i\ge 1$.
%So we consider another flow by changing $c_1$. When we take as
Then $c_1=-1/(2\mu\kappa^2)$. %, the initial value of $\tau_1$ is zero. 
%This case corresponds to the original RS model. 
By demanding also
$\tau_2(0)=0$, we obtain the following solutions,
\beq
 \tau_1={1\over 2\mu\kappa^2}\left(1-e^{2\mu y}\right),
\quad \tau_2={1\over 4\mu^3\kappa^2}\left(\mu y-e^{2\mu y}
   +{1\over 4}e^{4\mu y}+{3\over 4}\right).
\eeq
%%% Added
Here we assumed $\tilde{\tau}_2=0$ for  simplicity. 
In this case, we obtain
$$
  S^{\rm eff}_b = \int d^4x\sqrt{-\hat{g}}
      \left({1\over 4\mu\kappa^2}\hat{R}
      -{1\over 8\mu^3\kappa^2}\left(-e^{2\mu y}
   +{1\over 4}e^{4\mu y}+{3\over 4}\right)
    \left(\hat{R}_{\mu\nu}\hat{R}^{\mu\nu}-{1\over 3}\hat{R}^2\right)
      +\cdots \right)
$$
\beq
~~~~~~~~~~~~~~~~~~~~~~+S_{\rm CFT},
   \label{bactionRS}
\eeq
where we used the following notations,
$g_{\mu\nu}=A^2(y)\hat{g}_{\mu\nu}$ and $\hat{R}=R(\hat{g})$.
This result implies that the gravitational constant is independent
of $y$ and is equal to that given at $y=0$. Then the running given above
does not alter the physics observed on the brane as far as  the
Einstein gravity is concerned. 

%%%%%% Modified 10.27
As for the new  higher derivative term or the
Weyl term, the coefficient does not include the
power series of $y$ like the anomaly term. 
%%%%%%%%%%%%%%%%%%%%% modified 2nd
Terms like
$e^{2\mu y}$ and $e^{4\mu y}$ should be estimated by rewriting them as
$e^{2\mu (y-y_0)}$ and $e^{4\mu (y-y_0)}$ near the boundary. They are
not small, but the first Einstein term is in this case proportional to 
$A(y_0)^2=e^{-2\mu y_0}$ which is large for $y_0\to -\infty$. Then, the higher
derivative terms become negligible when $e^{2\mu (y-y_0)}/A(y_0)\ll 1$ and
$m^2/\mu^2 \ll 1$.
Under this condition, we can show
the irrelevant terms are also negligibly small near
the boundary.
%%%%%%%%%%%%%%%%%%%%% modified 2nd
It would be possible to remove the constant $3/4$ by modifying the counter 
term $S_{\rm CT}$, but the situation depends on the renormalization
condition of the effective action.
%%%%%%%%%%%%%%%%%%%%%%%%%%%%%%
%\footnote{When the brane position is taken at $y=y_0$, the warp factor is
%written as $A(y)=A(y_0)\sinh(\mu[y_H-|y-y_0|])$ and 
%$A(y_0)={\sqrt{\lambda}\over \mu}\sinh(\mu[y_H-y_0])$. And the brane starts
%from $y_0$ to the extended space direction $y>y_0$ according to the same
%flow equations given here.
%}.
%%%%%%%%% Modified
In any case, we arrive at the following result,
\beq
  S^{\rm eff}_b = \int d^4x\sqrt{-\hat{g}}\left(
      {A(y_0)^2\over 4\mu\kappa^2}\hat{R}+O(R^2)\right)
      +S_{\rm CFT},
   \label{EactionRS}
\eeq
%%%%%%%%%%%%  modified 2nd
%where the irrelevant terms are neglected by considering 
in the region of $y_0$ stated above.
%%%%%%%%%%%%%%%
We call this action  an invariant form under brane running since 
$A(y_0)^2\over 4\mu\kappa^2$  is independent of the running brane
position $y$. This implies that theories with different $\tilde{S}_b$
are considered to be the same when they are on the same flow line.
An interesting example for such $\tilde{S}_b$, 
which has the induced Einstein term in it, is 
the brane gravity model \cite{KTT,Dvali,CEHT}. 

\vspace{.3cm}
%%%%%%%%%%%%%% New Modified 2nd
From (\ref{EactionRS}),
the 4d Planck mass, $M_{\rm pl}^2=1/2\kappa_4^2$, is related to 
the 5d one, $M_5^3=1/2\kappa^2$, as 
\beq
M_{\rm pl}^2={M_5^3A(y_0)^2 \over 2\mu} \; .
\label{Mpl}
\eeq 
%We notice that %we can set $A(0)=1$ in (\ref{EactionRS}) for simplicity,
This condition implies a consequence that is of interest for  
phenomenological analysis:  by assuming  $M_5 \approx \mu$, 
we find $M_5 \ll M_{\rm pl}$ since $1\ll A(y_0)$. This might be 
an explanation of the hierarchy when the mass scale is given by $M_5$ for 
the field theory on the boundary.
%%%%%%%%%%%%%%%%%%%%%%
%%%%%%%%%%%%%%
%\vspace{.5cm}

In the next sub-section, we consider the case
of positive $\lambda$ and find a similar invariant form in spite of
the conformal non-invariance. However we give the discussion at
$y_0=0$ for  simplicity.

\vspace{.3cm}
\subsection{$\Lambda <0$ and $\lambda >0$}

Next is the case of $\Lambda <0$ and $\lambda >0$. 
Using $A(y)$ given in the previous section, the equation
for $\tau_1$ is solved as
\beq
  \tau_1={\sinh\left(2\mu(y_H-|y|)\right)+2\mu y-b_1\over
    4\mu\kappa^2\sinh^2\left(\mu(y_H-|y|)\right)},  \label{tau1-desitt}
\eeq
where $b_1$ is an integration constant. 

\vspace{.3cm}
In this case,
the conformal invariance is broken and we saw the flow of $\tau$ 
in section 3.2. However, $\tau$ of (\ref{desitt}) is considered to be
constructed by many $\tau_i$, and its
$y$-dependence is determined by the infinite number of $\tau_i$.
The same thing might be expected for $\tau_1$ given by
(\ref{tau1-desitt}).
Actually we can see that (\ref{tau1-desitt}) 
coincides with $\tau_1$ given in (\ref{sol-desitter})
in the limit of $\lambda=0$,
\beq
 \tau_1\rightarrow {1\over 2\mu\kappa^2} \label{limit}
\eeq
as expected. But in general, $\tau_1$ in (\ref{tau1-desitt}) 
deviates from the above value (\ref{limit}) for finite
$\lambda$.

\vspace{.5cm}
As in the previous subsection, we can estimate the effective brane
action by adjusting $b_1$ such that $\tau_1(0)=0$ as in the case
of $\lambda=0$. Then it is determined,
\beq
 b_1=\sinh(2\mu y_H)  \label{b1-bound}
%=2{\mu^2\over \lambda}\sqrt{1+{\lambda\over \mu^2}}.
\eeq
As for $S_{\rm CT}$, 
%we should modify the previous formula
%(\ref{bactionCT}) since $\tau_0$ and $\tau_1$ given for $\lambda>0$
%include many contributions coming from higher derivative terms.
for the $\lambda=0$ case, we can see that the first two terms in 
(\ref{bactionCT}) 
are obtained by substituting the classical solution into the bulk
action and integrating over $y$. Also for $\lambda>0$ case, we use this form
of counter term,
%Then $S_{\rm CT}$ consists of the classically induced 4d action and something, 
which cancels the UV divergent terms.
%$S_{\rm CT}$ obtained in this way includes
%the contributions from higher derivative terms through $A(y)$ since
%it is solved in terms of $\tau$ given in 3.2.

Replacing $S_{\rm CT}$
by the effective 4d action obtained by substituting classical
solution into the bulk action, we obtain
\beq
 S^{\rm eff}_b = \int d^4x\sqrt{-\hat{g}}%\left\{
        {1\over 2\kappa_4^2}\left(\hat{R}-
6\tilde{\lambda}+O(R^2)\right)+S_{\rm QFT}%+\cdots \right\},
  \label{effactiondS2}
\eeq
where 
\beq
 {1\over 2\kappa_4^2}={1\over 2\kappa^2}\int_0^{y_H}dy~ A^2(y), \quad
 \tilde{\lambda}=2\lambda\kappa_4^2 \int_y^{y_H}dy~ A^2(y).
\eeq
The corresponding field theory is replaced by quantum field theory
(QFT) since the bulk conformal symmetry would be broken.
%%%%%%%%%%%%%% Modified 12.3
We notice here that
the Planck constant $\kappa_4$ is exactly the same one with that given
at $y=0$ \cite{GY}. While the
effective cosmological constant $\tilde{\lambda}$ decreases with $y$ 
and it tends to zero at $y=y_H$.
On the other hand, the geometry of the 4d slice perpendicular to $y$
is charachterized by the scale factor, $a_0(t)=e^{\sqrt{\lambda}~t}$. 
This implies that the decreasing of $\tilde{\lambda}$ is compensated by
the higher derivative terms denoted by $O(R^2)$ in (\ref{effactiondS2})
\cite{Sta,Vilen}.
Then the equation (\ref{effactiondS2}) is well approximated by the first
two terms in the region of 
$\tilde{\lambda}\sim \lambda$ or $y\sim y_0=0$.

This implies that the inflation is
possible even if the cosmological constant is zero. 
In this case, however we should
not neglect the higher derivative terms, which appear 
when the brane position is changed.
Then, the higher derivative terms could be considered as a 
candidate of the dark
energy, which sustains the inflation, instead of the cosmological constant. 
This point we will make more clear when discussing
 the case of two branes in the next section.

\vspace{.5cm}

\section{Two parallel branes}

The system of two branes embedded in AdS$_5$ bulk 
is considered in this section. 
The action is given as  $S=S_5+S_{2b}$, where 
\beq
 S_{2b}=-\tau_{hid}
%\delta(y-y_{hid})
\int d^4x\sqrt{-g}
        -\tau_{vis}
%\delta(y-y_{vis})
\int d^4x\sqrt{-g} \; . 
\eeq 
The first (second) term represents the hidden (visible) brane 
with a plus (minus) tension $\tau_{hid}$ ($\tau_{vis}$) located at
$y=y_{hid}$ ($y=y_{vis} < y_H$). The bulk space between the two branes, 
$y_{hid} \le y \le y_{vis}$, is projected on an orbifold $S^1/Z_2$; 
namely, the bulk action $S_5$ is defined by 
(\ref{action}) in the compactified region. 
%\beq
%S_5=\int_{y}^{y_{vis}} dy L_5 \, .
%\eeq

The hidden brane is running 
from the original position $y=y_{hid}$ toward $y=y_{vis}$ 
with the brane running method 
shown in the previous sections. 
%%%%%%%%%%%%%%%%%%%%%%%%%%%
The tension $\tau_{hid}(y)$ of the hidden brane 
at a running position $y$ is then 
obtained by (\ref{tau0-lambda0}) for $\lambda=0$ and (\ref{desitt}) 
for $\lambda>0$. The visible brane, on the other hand, is  fixed 
at $y=y_{vis}$ with the tension 
%$\tau_{vis}$ which is obtained from $A(y)$:
\beq
 \tau_{vis}=-{6\mu\over \kappa^2}~~~ {\rm for}~~~\lambda=0 \; ,
\eeq
\beq
 \tau_{vis}=-{6\mu\over \kappa^2}\sqrt{1+{\lambda\over \mu^2A(y_{vis})^2}}~~~ 
         {\rm for}~~~\lambda>0 \; .
\eeq
%The visible brane does not run, so $\tau_{vis}$ is fixed as well as 
%the position of the brane. 

\vspace{.5cm}
Let us consider the hidden brane at a running position $y$. 
In the effective braneworld action, 
$S_{2b}^{\rm eff} \equiv S_{2b}/2+S_{\rm CT}+S_{\rm QFT}$, 
$S_{\rm CT}$ is obtained with the classical approximation, i.e. by substituting
the classical solution $A(y)$ into $S_5$, 
% as stated above, 
while quantum effects 
are included in $S_{\rm QFT}$. We then obtain 
the effective braneworld action for a running hidden brane:
% in which the hidden brane is 
% located at $y$ smaller than $y_{vis}$:
\beq
 S_{2b}^{\rm eff}={1\over 2}S_{2b}+S_5^{\rm classical}+S_{\rm QFT}
       \; ,
\eeq
%where $S_{b,hid}^{\rm eff}$ is given in the previous section for the
%running brane and 
\beq
S_{2b}= 
%S_{b,hid}=-{1\over 2}
%\delta(y-y_{hid})
- \int d^4x\sqrt{-g} \cdot \tau_{hid}(y)
- \int d^4x\sqrt{-g}\cdot \tau_{vis}
%          +\int_{y}^{y_H}L_5^{\rm classical} 
\; ,
\eeq
%\beq
%S_{b,vis}=-{1\over 2} 
%%\delta(y-y_{vis})
%\int d^4x\sqrt{-g}\cdot \tau_{vis}
%%          -\int_{y_{vis}}^{y_H}L_5^{\rm classical} 
%\; ,
%\eeq
where
\beq
  \tau_{hid}(y)=\tau_0+\tau_1 R
      +\tau_2\left(R_{\mu\nu}R^{\mu\nu}-{1\over 3}R^2\right)
      +\tilde{\tau}_2R^2+\cdots  \; ,
\eeq
and in the limit $y \to y_{hid}$, 
$\tau_i(y)$ tends to $\tau_{hid}$ for $i=0$ and 
vanishes for other $i$. 
The explicit form of $S_{2b}^{\rm eff}$ is obtained by calculating 
running parameters $\tau_{i}$ 
with the running brane method:
\beq
S_{2b}^{\rm eff}=\int d^4x \sqrt{-\hat{g}}
{1\over 2\kappa_4^2}
\left( \hat{R}
-6\tilde{\lambda} + O(R^2)
\right)
+S_{\rm QFT}
\eeq
with 
\beq
   {1\over 2\kappa_4^2}={1\over 2\kappa^2}\int_{y_{hid}}^{y_{vis}} dy A^2(y)
\; ,  \quad  \quad
\tilde{\lambda}=\lambda {\kappa_4^2 \over \kappa^2} \int_y^{y_{vis}}dy~ A^2(y)
%  {1\over 2\bar{\kappa}_4^2}={1\over 2\kappa^2}\int_{y}^{y_{vis}} dy A^2(y) 
\; .
\eeq
This  effective action is obviously independent of $y$ 
in its first term.  In general, on the other hand, the $\tilde{\lambda}$ term, 
the higher derivative terms and $S_{\rm QFT}$ depend on $y$. 
When the two branes are near the boundary, 
as already argued in the one-brane case, 
all the $y$ dependent terms are suppressed for small $m/\mu$. 
The effective action surely describes an inflation in 4d spacetime 
with the net effect of $\tilde{\lambda}$ and the higher derivative terms, 
since the scale factor of the 4d spacetime is $a_0(t)=e^{\sqrt{\lambda}~t}$, 
independently of the position of the hidden brane.

\vspace{.5cm}
In general, it is quite hard to calculate $S_{2b}^{\rm eff}$, 
since so is  $S_{\rm QFT}$. This problem is circumvented by the following 
prescription. 
When the hidden brane is placed upon the visible one, 
i.e. at $y=y_{vis}$, there is no bulk, so $S_{\rm CT}$ and $S_{\rm QFT}$
vanish. Then,  $S_{2b}^{\rm eff,ren}$ is reduced to a calculable quantity, 
$S_{2b}/2$. 
Thus, we can evaluate $S_{2b}^{\rm eff}$ by 
a hidden-brane running 
from the original position $y=y_{hid}$ to $y=y_{vis}$ \cite{LMS}

\vspace{.5cm}
The expectation value of $-S^{\rm eff}_{2b}$ with the background field 
$a_0(t)$ can be regarded as the effective potential $V_{\rm eff}$ 
for the brane positions, $y_{hid}$ and $y_{vis}$. 
Now our discussion is focused on the case of small $\lambda$ and
the brane running near the boundary. 
In this situation, 
the higher derivative terms, at least of order $\lambda^2$, are negligible 
compared with the leading term $R$ of order $\lambda$. 
For such small $\lambda$, 
the effective potential is given in the simple form 
\beq
%   V_{\rm eff}(y_{hid},y_{vis})= - a_0(t)^3 {3 \lambda \over \kappa^2}
%\int_{y_{hid}}^{y_{vis}} dy A^2(y)
   V_{\rm eff}(y_{hid},y_{vis})= - a_0(t)^3 {6 \lambda \over \kappa_4^2}\; .
\label{Veff}
\eeq
The potential becomes small as the interbrane distance, $d=y_{vis}-y_{hid}$, 
increases; note $\hat{R}=12\lambda$ here and
that this statement is valid for small $d$, 
since we consider the two-brane system near the boundary. 
This indicates that the two-brane system located near the boundary 
is unstable. This conclusion might be compatible 
with the radion stabilization analysis \cite{BDL}.

\vspace{.5cm}
As for $\lambda=0$, $V_{\rm eff}$ vanishes for any value of $d$, 
so $d$ is undetermined. 
The distance can be stabilized by introducing a massive bulk scalar \cite{GW}. 
The Goldberger-Wise stabilization mechanism is seen 
through the effective potential 
which has a minimum at a finite distance. 
As for a small and finite $\lambda$, however, (\ref{Veff}) should be add to 
the Goldberger-Wise stabilization potential. 
There is a possibility that 
the net potential has no minimum, when (\ref{Veff}) 
is larger than the stabilization potential in magnitude. 

\vspace{.5cm}
From a cosmological viewpoint, regarding 
%As a general remark on 
the physics of the two-brane setting, let us finally point out how the two parallel branes are interconnected, %even 
in the case of a fixed separation distance. This is a consequence of Einstein's equations in the bulk, together with the junction conditions on the branes. Let us put $y_{hid} =0$. Then, with $\lambda_0=\kappa^2 \tau_0^2/36+\Lambda/6$ and similarly $\lambda_{vis}=\kappa^2\tau_{vis}^2/36+\Lambda/6$ we obtain in the static case, taking the junction conditions
\begin{equation}
\frac{A'}{A}\Bigg|_0=-\frac{\kappa^2}{6}\tau_0, \quad \frac{A'}{A}\Bigg|_{vis}=\frac{\kappa^2}{6}\tau_{vis} 
\end{equation}
into account, the following relationship \cite{brevik03}:
\begin{equation}
\lambda_{vis}A^2(y_{vis})=\lambda_0.
\end{equation}
Since $A(y)<1$ when $y>0$ according to (\ref{desit0}) (here assuming $A(0)=1$), we see that
\begin{equation}
\lambda_{vis}>\lambda_0,
\end{equation}
which in turn means that $\tau_{vis}^2>\tau_0^2$. As the effective four-dimensional cosmological constant $\lambda$ is equivalent to the presence of an isotropic fluid on the brane obeying the equation of state $p=-\rho$ for a `vacuum' fluid \cite{brevik03}, we see that the magnitude of the energy density is largest on the visible brane, $|\rho_{vis}|>|\rho_0|$. When the position of the hidden brane is running, the disturbance that this brane is subject to, has to influence the other brane also.

%%%%%%%%%%%%%%%%%%%%%%%%%%%

%%%%%%%%%%%%%%%%  Conclusion %%%%%%%%%%%%%%%%%%
\section{Concluding remarks}

We have studied the effective braneworld action derived by brane running, for 
one-brane and two-brane models with and without a 4d cosmological constant 
$\lambda$ on the brane.  
For the one brane model, the effective action is regularized such that it is 
finite at the boundary. 
In the limit $\lambda \to 0$, we can see %in particular, 
the quadratic-curvature term, %has an UV divergence 
which can be regarded as the anomaly 
produced by the one-loop  CFT (the  AdS/CFT correspondence),
in the action of the running brane. 
This term is however divergent at the boundary, 
%$y\to -\infty$, 
so it must be cancelled by some regularization procedure.
We give here such a regularization scheme. 
In the resultant effective braneworld action, 
the effective 4d Planck mass is 
invariant under the brane running for any value of $\lambda$. 
In the case of finite $\lambda$, interestingly, 
the $\lambda$ term diminishes and finally vanishes as the brane position 
inceases toward the horizon, while 
higher-derivative terms, and a cutoff QFT term, increase. 
This means that the inflation is realized by the higher-derivative and 
cutoff QFT terms even if $\lambda=0$, since the scale factor on the brane 
is always $a_0(t)=\exp{(\sqrt{\lambda}t)}$ during brane running. 

   When we consider the brane running near the boundary and 
the KK modes of small $m^2$ relative to $\mu^2$, 
the higher-derivative and cutoff QFT terms are suppressed, and 
the $\lambda$ term is almost constant. 
Thus, the brane running gives the same low-frequency physics 
at any brane position, as long as the conditions mentioned above 
are satisfied. The running brane action includes an "induced"
gravity, since $\tau_1$ generally becomes finite as a result of brane running. 
Then it is not essential whether one takes into account the Einstein term in the brane action or not.

\vspace{.5cm}
It is hard to derive the definite form of $S_{\rm QFT}$ in 
the effective braneworld action. However,
for the two brane model, we can get 
the effective braneworld action without $S_{\rm QFT}$, 
%, that is, 
%with the hidden-brane running from 
by moving the hidden-brane from the original position to 
the visible bane, since $S_{\rm QFT}$ finally 
vanishes due to the definition.
Applying the action principle 
to the effective braneworld action thus calculated, 
we  see that 
the two-brane system located near the boundary 
is unstable. %stable for the case of small but finite $\lambda$. 
This result might be 
compatible with the analysis of the radion stabilization 
\cite{BDL}. 
In the limit $\lambda \to 0$, %on the other hand, 
the vacuum value of the effective braneworld action vanishes, 
so the separation between two branes is undetermined as expected
from the restored conformal invariance.

\vspace{.3cm}
We considered only the gravitational part, but we could extend
the same analysis to the case where other fields are existing. Especially
bulk scalars, which couple to the brane, will play an important role
in the determination of distance between two branes in the RS I model.
Another important problem to be studied is the estimate of massive
KK modes in the effective braneworld action. They are essential for
the corrections of  Newton's law and for some cosmological problems.
We intend to discuss these points in a future paper.
%     As an interesting application of the brane running, we
%consider the one-brane system in 
%which the 5d parameters such as $M_5$ and $\mu$ are much smaller than 
%the 4d Planck mass. The large difference between the 4d Planck mass and 
%the 5d parameters are understood by introducing a large warp factor $A(0)$ 
%at the brane position. 
%The RS model II with large $A(0)$ is reduced to the 
%cosmological model with brane gravity \cite{Dvali}, when a UV cutoff is moved 
%down from the original position $y=0$ to $y=y_0 > 0$ 
%which satisfies $A(y_0)=1$. 
%Thus, both the models yield the same low-frequency physics on 4d spacetime. 
%The cosmological model with brane gravity is then a low-energy effective 
%theory of the RS model II with a large warp factor at the brane position. 

%%%%%%%%%%%%%%%%%%%%%%%%%%%%%%%%
\section*{Acknowledgments}
This work has been supported in part by the Grants-in-Aid for
Scientific Research (13135223, 14540271)
of the Ministry of Education, Science, Sports, and Culture of Japan.


\begin{thebibliography}{99}

\bibitem{M1} J.~Maldacena, Adv. Theor. Math. Phys. {\bf 2} (1998) 231, 
        ({\tt hep-th/9711200}).
\bibitem{GKP1} S.S. Gubser, I.R. Klebanov, and A.M. Polyakov Phys. Lett.
        {\bf B 428} (1998) 105, ({\tt hep-th/9802109}).
\bibitem{W1} E.~Witten, Adv. Theor. Math. Phys. {\bf 2} (1998) 253, 
        ({\tt hep-th/9802150}).
\bibitem{Poly1} A.M. Polyakov, Int. J. Mod. Phys. {\bf A14} (1999) 645,
        ({\tt hep-th/9809057}).
\bibitem{RS1} L. Randall and R. Sundrum, Phys. Rev. Lett. {\bf 83} (1999)
3370,
        ({\tt hep-ph/9905221}).
\bibitem{RS2} L. Randall and R. Sundrum, Phys. Rev. Lett. {\bf 83} (1999)
4690,
        ({\tt hep-th/9906064}).
\bibitem{LMS} A. Lewandowski, M. May and R. Sundrum, Phys. Rev. {\bf D67}
(2003) 024036 ({\tt hep-th/0209050}). 
\bibitem{LR} A. Lewandowski, and M. Redi, ({\tt hep-th/0305013}). 
\bibitem{Re} M. Redi, JHEP 0305:032, 2003({\tt hep-th/0304014}). 
\bibitem{U} N. Uekusa, ({\tt hep-th/0307107}). 
\bibitem{DL} M.J.~Duff and J.T. Liu, Phys. Rev. Lett. {\bf 85} (2000) 2052, 
        ({\tt hep-th/0003237}).
\bibitem{NOZ} S. Nojiri,S. Odintsov and S. Zerbini, 
Phys. Rev. D62 (2000) 064006 ({\tt hep-th/0001192}).
\bibitem{GR}
    W.D. Goldberger, and I.Z. Rothstein, Phys. Lett. B 491(2000)339, 
      ({\tt hep-th/0007065}). 
\bibitem{GGH} H. Georgi, A.K. Grant, G. Hailu, Phys.Lett. B506 (2001) 207-214,
        ({\tt hep-ph/0012379}).
\bibitem{MOZ}  K.A. Milton, S.D. Odintsov, S. Zerbini, 
      Phys. Rev. D65 (2002) 065012 ({\tt hep-th/0110051}).
\bibitem{bre}
I. Brevik, K. Ghoroku, S. D. Odintsov and M. Yahiro, Phys. Rev. {\bf 66}
   (2002) 064016, ({\tt hep-th/0204066}). 
\bibitem{Gidd} S.B. Giddings, E. Katz and L. Randall, JHEP {\bf 03} (2000) 023,
({\tt hep-th/0009176}).
\bibitem{GKa} S.B. Giddings and E. Katz, 
 ({\tt hep-th/0002091}).
\bibitem{HeSk} M. Henningson and K. Skenderis, 
        JHEP {\bf 07} (1998) 023, ({\tt hep-th/9806087}).
\bibitem{GY}
      K. Ghoroku, and M. Yahiro, Phys. Rev. {\bf 66}
   (2002) 124020, ({\tt hep-th/0206128}). 
\bibitem{Sta}
    A.A. Starobinsky, Phys. lett. 91B(1980), 99. 
\bibitem{Vilen}
    A. Vilenkin, Phys. Rev. D32(1985), 2511. 
\bibitem{BDL} P. Binetruy, C. Deffayet and D. Langlois, 
         Nucl. Phys. {\bf B615} (2001) 219. 
\bibitem{GW}
    W.D. Goldberger, and M.B. Wise, Phys. Rev. Letters 83(1999), 4922. 
\bibitem{KTT} E. Kiritisis, N. Tetradis, T.N. Tomaras,
          ({\tt hep-th/0202037}). 
\bibitem{Dvali}
    G. Dvali, G. Gabadadze and M. Porrati, Phys. Letters B485 (2000), 208. \\
    G. Dvali and G. Gabadadze, Phys. Rev. D63(2001), 065007. 
\bibitem{CEHT} C. Csaki, J. Erlich, T.J. Hollowood and J. Terning,
Phys. Rev. {\bf 63} (2001) 065019, ({\tt hep-th/0003076}). 
\bibitem{brevik03}
    I. Brevik, K. B{\o}rkje and J. P. Morten, gr-qc/0310103.



%%%%%%%%%%%%%%%%%%%%%%%%%%%%%%%%%%%%%%%%%%%%%%%%%%%%%%%%%%%%%%%%%%%%%%%%%%
\end{thebibliography}
\end{document}